\documentclass[12pt]{iopart}
\usepackage{graphicx}
\usepackage{epsfig}
\usepackage{latexsym}
\usepackage{iopams}

\begin{document}

\title[Shielding efficiency and $E$($J$) characteristics measured on Bi-2212 cylinders]{Shielding efficiency and $E$($J$) characteristics measured on large melt cast Bi-2212 hollow cylinders in axial magnetic fields}

\author{J-F Fagnard$^1$,~S Elschner$^{2, 3}$,~J Bock$^3$,~M Dirickx$^1$,~B Vanderheyden$^4$,~and P Vanderbemden$^4$}
\address{$^1$ SUPRATECS, CISS Dept., Royal Military Academy, Belgium}
\address{$^2$ University of Applied Science Mannheim, D-68163 Mannheim, Germany}
\address{$^3$ Nexans SuperConductors GmbH, D-50351 H\"{u}rth, Germany}
\address{$^4$ SUPRATECS, Dept. of Electrical Engineering and Computer Science (B28), University of Li\`ege, Belgium}
\ead{fagnard@montefiore.ulg.ac.be}

\begin{abstract}

We show that tubes of melt cast Bi-2212 used as current leads for LTS magnets can also act as efficient magnetic shields. The magnetic screening properties under an axial DC magnetic field are characterized at several temperatures below the liquid nitrogen temperature ($77~\mathrm{K}$). Two main shielding properties are studied and compared with those of Bi-2223, a material that has been considered in the past for bulk magnetic shields. The first property is related to the maximum magnetic flux density that can be screened, $B_{\mathrm{lim}}$; it is defined as the applied magnetic flux density below which the field attenuation measured at the centre of the shield exceeds 1000. For a cylinder of Bi-2212 with a  wall thickness of $5~\mathrm{mm}$ and a large ratio of length over radius, $B_{\mathrm{lim}}$ is evaluated to $1~\mathrm{T}$ at $T = 10~\mathrm{K}$. This value largely exceeds the $B_{\mathrm{lim}}$ value measured at the same temperature on similar tubes of Bi-2223. The second shielding property that is characterized is the dependence of $B_{\mathrm{lim}}$ with respect to variations of the sweep rate of the applied field, $\mathrm{d}B_{\mathrm{app}}/\mathrm{d}t$. This dependence is  interpreted in terms of the power law $E = E_\mathrm{c}(J/J_\mathrm{c})^n$ and allows us to determine the exponent $n$ of this $E$($J$) characteristics for Bi-2212. The characterization of the magnetic field relaxation involves very small values of the electric field. This gives us the opportunity to experimentally determine the $E$($J$) law in an unexplored region of small electric fields. Combining these results with transport and AC shielding measurements, we construct a piecewise $E$($J$) law that spans over 8 orders of magnitude of the electric field.
\end{abstract}
\submitto{\SUST}
\maketitle


\section{Introduction} \label{s:introduction}

Bulk tubular samples of high temperature superconductor (\emph{HTS}) are commonly commercialized as current leads for low temperature superconducting magnets \cite{Herrmann1993} or used as a component for fault current limiters (\emph{FCL}) \cite{Dommerque2010} and cryogenic current comparators (\emph{CCC}) \cite{Grohmann1974, Giunchi2007}. In addition, advantage can be taken of the tubular form in order to use the superconducting material as a magnetic screen \cite{Ohta1999,Denis2007a,Fagnard2009a}. Several HTS materials, including MgB$_2$ or Bi- and Y-based cuprates, are candidates for building bulk hollow cylinders. Amongst the different techniques, the melt casting process (\emph{MCP}) has proved as an efficient method for manufacturing Bi$_2$Sr$_2$CaCu$_2$O$_8$ (Bi-2212) of arbitrary shape \cite{Bock1995}. In the particular case of tubes, the process involves a complete melting of the superconducting compound which is then cast into moulds of desired dimensions using centrifugal technique. The samples synthesised in this way are used as core of \emph{FCL} elements and current leads \cite{Herrmann1993, Dommerque2010, Bock1995}. In the present work we characterize in detail the ability of such samples to act as a magnetic screen.

Basically, characterizing a magnetic shield requires, in priority, the knowledge of the maximum magnetic field that can be shielded with a given attenuation level. This maximum field is linked to the maximum shielding currents that can flow in the superconductor and thus to the critical current density, $J_\mathrm{c}$, of the superconducting material. The field and temperature dependence of $J_\mathrm{c}$ is often determined through transport or magnetic measurements on small samples cut from the magnetic shield. In the present work we aim at characterizing the properties of whole, large tubular samples (length $80~\mathrm{mm}$, external radius $26~\mathrm{mm}$) by using a Hall probe sensor placed in the centre of the shield. In addition to being non-destructive, the method has the following advantages. First, the true shielding properties of the screen are determined since possible inhomogeneities of the critical current are integrated, which cannot be detected from measurement on a small sample. Second, the measurements by a sensor placed at the centre of the screen are usually much less sensitive to geometric (demagnetization) effects than measurements of the whole sample magnetization, using e.g. a SQUID or a VSM. Third, the magnetic induction inside the cylinder is influenced only by macroscopic circular current loops flowing in the walls of the screen, i.e. \emph{intergranular} currents; this is in contrast with the magnetic moment on a plain sample which is possibly caused by \emph{intergranular} as well as \emph{intragranular} currents \cite{Muller1999, Fagnard2002}.

The performance of magnetic shields made of various high-temperature superconductors (YBCO, Bi-2212, Pb-doped Bi-2223 and $\mathrm{MgB}_2$) has been demonstrated by several authors for nearly two decades \cite{Miller1993, Plechacek1996, Cavallin2006}. These investigations on the magnetic properties of whole superconducting screens are usually carried out at \emph{fixed} temperatures, i.e. liquid helium or liquid nitrogen temperatures. Reports of the temperature dependence of the magnetic shielding performance are scarce and involve superconductors of small dimensions \cite{Fournier1994}. To our knowledge however, the temperature dependence of the magnetic shielding properties of whole, large Bi-2212 hollow cylinders has not been studied yet. This temperature dependence will be investigated in the present paper.

Superconductors are relevant for shielding low frequency - or even DC - magnetic fields. This is because high frequency magnetic fields can be easily shielded by any normal metal through the skin effect, in practice above 1 kHz. In the low frequency regime, the efficiency of the magnetic shielding of high-temperature superconductors may significantly depend on the sweep rate (or frequency in AC) of the applied magnetic field. This behaviour arises from magnetic relaxation effects, which are strongly linked to the non-linear relationship between the shielding currents, $J$, and the electric field, $E$, usually in the form of a power law $E \sim J^n$. A precise knowledge of the $E$($J$) curve of a superconductor is therefore of importance in order to determine how the magnetic shielding properties are influenced by the frequency of the applied magnetic field. In addition, it is often extremely instructive to combine transport and magnetic measurements on the same material, since the electric field induced in low frequency magnetic measurements is usually much smaller than the electric field produced by the voltage drop across a sample subjected to a transport current \cite{Caplin1994, Vanderbemden2002}. Studies of small samples was performed in order to determine the $E$($J$) characteristics of several superconductors \cite{Ries1992, Paul1991, Yamasaki2000}. In the present work, we propose to combine magnetic and transport measurements to determine the $E$($J$) characteristic on large Bi-2212 tubes. 

Our paper is organized as follows. First, the threshold - or ``limiting" - magnetic induction below which the shielding occurs, $B_{\mathrm{lim}}$, is measured between $10~\mathrm{K}$ and $77~\mathrm{K}$, this temperature dependence of $B_{\mathrm{lim}}$ is compared to the behaviour of similar superconducting hollow cylinders made of Bi-2223 that were studied in our previous works \cite{Denis2007a, Fagnard2009a, Denis2007b, Fagnard2009b}. Next, in order to obtain information about the non-linear $E$($J$) relationship, the dependence of $B_{\mathrm{lim}}$ on the sweep rate of the applied magnetic field is determined and analysed. The time relaxation of the shielding currents is also studied to explore a window of electric field that covers several orders of magnitude. These results are completed with AC shielding measurements and transport four-point measurements on the same bulk Bi-2212 tubes in order to build a piecewise $E$($J$) graph that extends over 8 decades.

\section{Experiment}\label{s:samples}

\begin{table}
\caption{\label{t:sampleBi2212}Geometrical characteristics of the Bi-2212 hollow cylinder.}
  \begin{indented}
   \item[]\begin{tabular}{@{}ll}
      \br
      Material & Bi$_{2}$Sr$_2$CaCu$_2$O$_8$ \\
      Length & $\ell  =80~\mathrm{mm}$ \\
      Inner radius & $a_1=8~\mathrm{mm}$\\
      Wall thickness & $e= 5~\mathrm{mm}$\\
      \br
    \end{tabular}
   \end{indented}
\end{table}

The studied sample is a hollow cylinder of Bi$_2$Sr$_2$CaCu$_2$O$_8$ produced by the melt casting process. Details on the preparation are discussed in \cite{Bock1995} and dimensions of the sample are shown in Table~\ref{t:sampleBi2212}. 
A microstructural study of the cylinder cross section reveals the existence of three concentric cylindrical zones \cite{Bock1995}.

\begin{figure}
\center
\includegraphics[width=8cm]{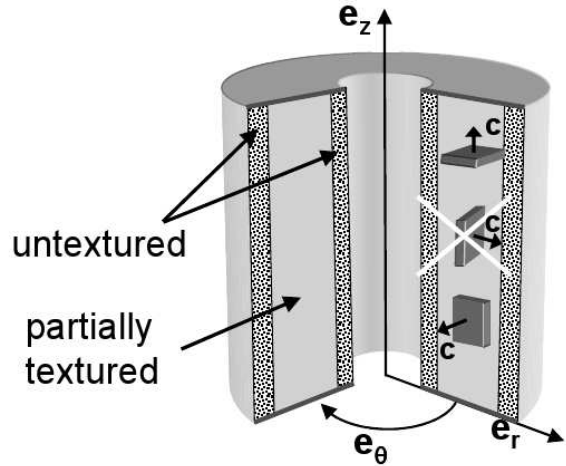}\caption
  {Schematic illustration of the microstructure of the melt cast Bi-2212 cylinder. A partially textured zone is surrounded by untextured zones at the inner and the outer surfaces of the cylinder. The microstructure reveals that, in the partially textured zone, the \emph{c}-axis of the platelets are rarely oriented parallel to the radial axis $\mathbf{e_r}$, the thermal gradient direction in the melt casting process. The \emph{c}-axis are distributed randomly along the axial direction, $\mathbf{e_z}$, and the azimuthal direction, $\mathbf{e_\theta}$.}\label{f:microstructure}
\end{figure}

The zone in the bulk of the wall of the tube, i. e. far from the inner and outer surfaces, exhibits a structure of bundles ($300$ to $500~\mathrm{\mu m}$) of platelets ($30$ to $50~\mathrm{\mu m}$) presenting low angle grain boundaries within the bundles. Because of the radial direction of the thermal gradient during the synthesis process, the \emph{c}-axis of the platelets are mostly perpendicular to the radial axis, $\mathbf{e_r}$, as shown schematically in \ref{f:microstructure}. The grains with the \emph{c}-axis parallel to the tube axis $\mathbf{e_z}$ or parallel to the azimuthal direction $\mathbf{e_\theta}$ (and all in between) occur with the same probability, whereas grains with the \emph{c}-axis parallel to $\mathbf{e_r}$ are rare. The middle zone thus has a non-uniform distribution of grain orientations, i.e. a partial texture. This zone is embraced by two nearly untextured areas. The outer part of the cylinder (typical thickness of $200~\mathrm{\mu m}$) is a glassy layer of small grains resulting from the rapid cooling on the mould walls. The inner zone is less dense because of the shrinkage inherent to the cooling in the melt casting process and does not present any particular texture either. Recent investigations on correlations between microstructure and superconducting properties have revealed that the partially textured zone has by far the largest critical current density \cite{Rikel2010}.

The measurements of the local magnetic induction at the centre of the sample at fixed temperatures between $10~\mathrm{K}$ and $77~\mathrm{K}$ are carried out in a Physical Property Measurement System (Quantum Design - PPMS). In the instrumental chamber where the sample is placed, a high sensitivity Hall sensor (Arepoc AXIS-3S) probes the local magnetic induction, $B_\mathrm{in}$, at the centre of a tubular sample subjected to an applied axial magnetic field ramped at a given sweep rate between $0.5~\mathrm{\mu T/s}$ and $1~\mathrm{mT/s}$. The temperature and the applied magnetic field, measured by the PPMS, as well as the Hall voltage, measured by a high sensitivity voltmeter (HP34420A), are monitored with a Labview\textregistered~interface.

A second set-up is also used for the characterization in a liquid nitrogen bath ($T = 77~\mathrm{K}$). This system involves a high sensitivity Hall probe (Arepoc - HHP-MP). The applied axial magnetic field is provided by a copper coil fed by a HP6030A DC power supply. The applied magnetic field at the centre of the coil can reach $\mu_0H_\mathrm{app} = 36~\mathrm{mT}$. The field is ramped up linearly, at sweep rates in the range [$5~\mathrm{\mu T/s}$, $20~\mathrm{mT/s}$], with a Labview\textregistered~interface that monitors simultaneously the Hall probe voltage sampled by a PCI-6281 National Instrument Data Acquisition Card.
Two concentric mu-metal ferromagnetic enclosures ensure the protection against stray magnetic fields. 

The transport measurements (DC) are carried out by using the common four-point technique. With a Keithley nV-meter, we measured the voltage for an increasing current, $I$, up to 2000 A. Considering the adequate geometrical dimensions (cross section $A = 3.30~\mathrm{cm}^2$), we obtain directly $E(J)$. The tube shaped samples are equipped with silver metal sheet contacts already integrated during the melt casting process \cite{Elschner1992}. The contact resistances are nearly independent of current with values of $0.2 - 0.5~\mathrm{\mu \Omega}$. The transport experiments are carried out first, for them the ends of the tubes with the metal sheet contacts are cut and magnetic experiments are performed.

The AC-shielding is measured (for details, see \cite{Elschner1999}), at a frequency of $50~\mathrm{Hz}$ with the sample mounted coaxially in a narrow coil (length $L=200~\mathrm{mm}$, number of turns $N=400$, gap between sample and coil about $1.5~\mathrm{mm}$). Coil and sample are immersed in liquid nitrogen at $T = 77~\mathrm{K}$. The sinusoidal AC-current through the coil can be adjusted between $0 - 60~\mathrm{A_{rms}}$. In the centre of the sample a pick-up coil (radius $R_{\mathrm{pu}}=5~\mathrm{mm}$, number of turns $N_{\mathrm{pu}}=100$) is placed and its waveform is recorded and averaged by summation.

\section{Results}
\label{s:results}
\subsection{Influence of the temperature on the shielding efficiency}
\label{ss:temperature}

\begin{figure}
\center
\includegraphics[width=8cm]{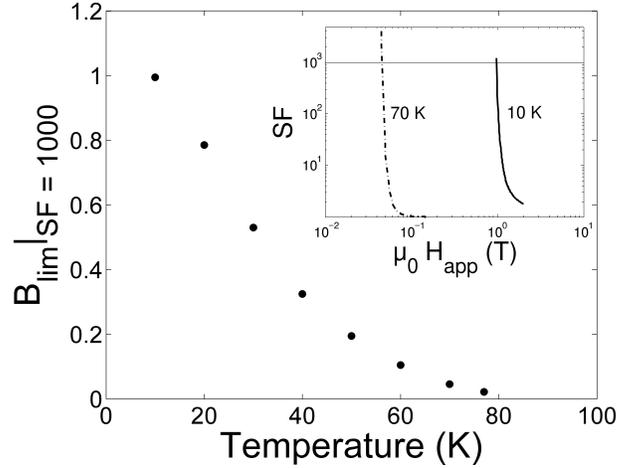}\caption
  {Temperature dependence of $B_{\mathrm{lim}}$, the applied magnetic induction that corresponds to a shielding factor equal to 1000. The data refer to measurements carried out at $\mu_\mathrm{0} \mathrm{d}H_{\mathrm{app}} /\mathrm{d}t = 125~\mathrm{\mu T/s}$. Inset: Shielding factor (\emph{SF}) measured at the centre of the cylinder at $10~\mathrm{K}$ and $70~\mathrm{K}$}\label{f:Blim_T}
\end{figure}

First, we study the maximum field that can be shielded at several fixed temperatures. At a given temperature between $10~\mathrm{K}$ and $77~\mathrm{K}$, the magnetic field, $H_{\mathrm{app}}$, is ramped up at a constant low sweep rate of $125~\mathrm{\mu T/s}$. The shielding factor, $SF = \mu_\mathrm{0} H_{\mathrm{app}}/B_{\mathrm{in}}$, a ratio of the applied magnetic flux density over the induction inside the cylinder, is calculated. We define the threshold magnetic induction, $B_{\mathrm{lim}}$, as the applied magnetic induction that corresponds to $SF=1000$. Above $B_{\mathrm{lim}}$, the magnetic field inside the cylinder is thus attenuated by less than 1000. The temperature dependence of $B_{\mathrm{lim}}$ is presented in figure \ref{f:Blim_T}. The inset of figure \ref{f:Blim_T} shows the applied field dependence of the shielding factor (\emph{SF}) measured at the centre of the cylinder for $T = 10~\mathrm{K}$ and $T = 70~\mathrm{K}$. The corresponding values of $B_\mathrm{lim}$ are indicated on the graph.

\begin{figure}
\center
\includegraphics[width=8cm]{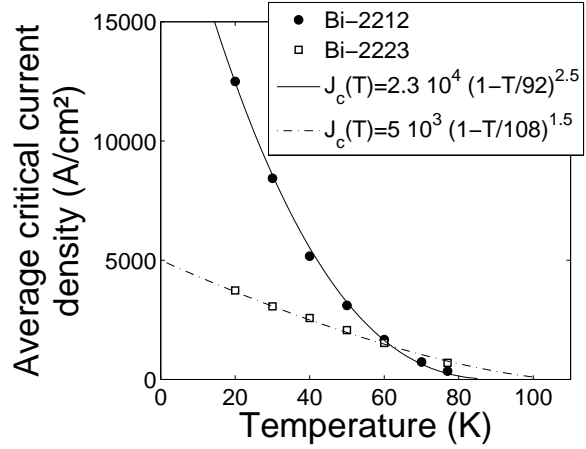}\caption
  {Temperature dependence of $J_{\mathrm{c}}$ calculated from shielding measurements of Bi-2212 (\fullcircle) and Bi-2223 hollow cylinders (\opensquare). The solid line presents a $J_\mathrm{c}(T)$ fit for Bi-2212 ($T_\mathrm{c}=92~\mathrm{K}$) and the dashed line presents a $J_\mathrm{c}(T)$ fit for Bi-2223 ($T_\mathrm{c}=108~\mathrm{K}$).}\label{f:compareJc}
\end{figure}

In view of comparing the temperature dependence of the magnetic shielding efficiency presented in figure \ref{f:Blim_T} to that of another material, we calculate the ratio, $J_\mathrm{c} = B_{\mathrm{lim}} / \mu_\mathrm{0} e$, between ($B_{\mathrm{lim}}/\mu_0$) and the thickness, \emph{e}, of the cylinder similarly to what is done in the Bean model with constant $J_\mathrm{c}$ \cite{Bean1962}. This ratio corresponds to an \emph{average} critical current density, $J_\mathrm{c}$,  in the superconducting wall and neglects the field dependence of $J_\mathrm{c}$. Figure~\ref{f:compareJc} presents the comparison between the average critical current densities of the Bi-2212 hollow cylinder and a tubular polycrystalline sample of Pb-doped Bi-2223 \cite{Fagnard2009a} whose physical characteristics are shown in Table~\ref{t:sampleBi2223}. The $J_c$ data plotted in figure \ref{f:compareJc} cover a temperature range between $20~\mathrm{K}$ and $77~\mathrm{K}$. Above $60~\mathrm{K}$, the average shielding currents flowing in the Bi-2223 material are larger than those of Bi-2212, e.g. at $77~\mathrm{K}$, the Bi-2223 material is characterized by a current density of $730~\mathrm{A/cm^2}$ whereas that determined for the Bi-2212 sample equals $350~\mathrm{A/cm^2}$. Note that the $J_\mathrm{c}$ values for both materials are determined  from magnetic shielding measurements on macroscopic tubular samples and refer thus to the \emph{intergranular} critical current density. Below $60~\mathrm{K}$, the shielding currents flowing in the Bi-2212 material are stronger than those of Bi-2223, e.g. at $20~\mathrm{K}$, the current density in Bi-2212 is 3 times that of Bi-2223. Knowing the respective critical temperatures of the materials, a usual law, which will be discussed in section \ref{s:discussion}, is used to fit the experimental data.

\begin{table}
\caption{\label{t:sampleBi2223} Geometrical characteristics of the Bi-2223 hollow cylinder.}
 \begin{indented}
   \item[]\begin{tabular}{@{}ll}
      \br  
      Material & Bi$_{1.8}$Pb$_{0.26}$Sr$_2$Ca$_2$Cu$_3$O$_{10+x}$ \\
      Length & $\ell  =80~\mathrm{mm}$ \\
      Inner radius & $a_1=6~\mathrm{mm}$\\
      Wall thickness & $e= 1.6~\mathrm{mm}$\\
      \br  
    \end{tabular}
   \end{indented}
\end{table}

\subsection{Influence of the sweep rate on the shielding efficiency} \label{ss:sweep rate}

\begin{figure}
\center
\includegraphics[width=8cm]{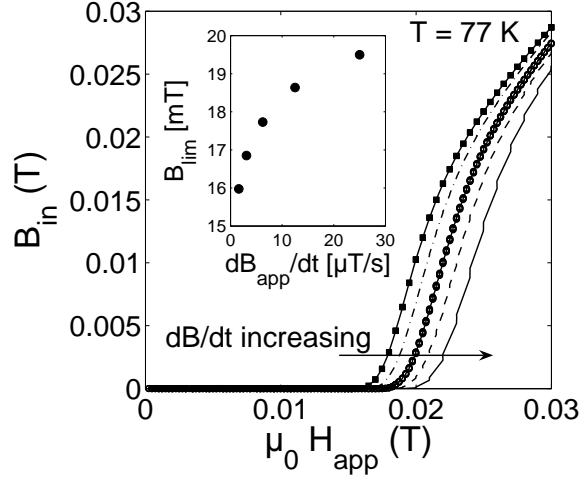}\caption
  {Sweep rate dependence of $B_{\mathrm{in}}(\mu_\mathrm{0}H_{\mathrm{app}})$ measured on Bi-2212 hollow cylinder at $77~\mathrm{K}$. Inset: Sweep rate dependence of $B_{\mathrm{lim}}$.}\label{f:dBdt}
\end{figure}

\begin{figure}
\center
\includegraphics[width=8cm]{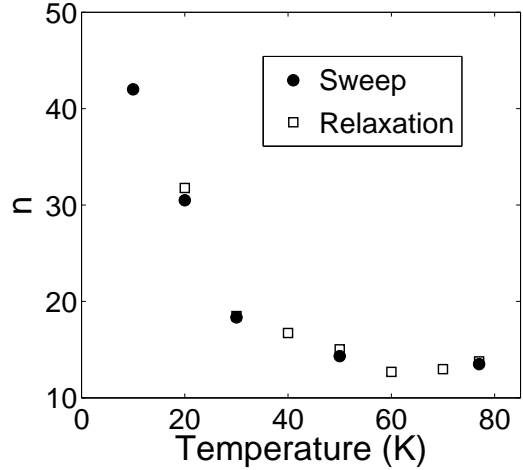}\caption
  {Temperature dependence of the exponent of the $\mathrm{d}B_{\mathrm{app}} /\mathrm{d}t(B_\mathrm{lim})$ power law measured on Bi-2212 hollow cylinder from sweep creep experiments (\fullcircle) and relaxation creep experiments (\opensquare).}\label{f:n_T}
\end{figure}

At a given temperature, $T=77~\mathrm{K}$, the local magnetic induction at the centre of the Bi-2212 hollow cylinder is measured for several sweep rates of the increasing magnetic field, the procedure is referred to as a \emph{sweep creep experiment} \cite{Caplin1994}. From the $B_{\mathrm{in}}(\mu_\mathrm{0}H_{\mathrm{app}})$ curves presented in figure \ref{f:dBdt}, we determine the threshold magnetic induction, $B_{\mathrm{lim}}$, corresponding to each sweep rate. From Faraday's law the electric field, $E$, at the mean radius of the cylinder is related to the time derivative of the magnetic flux and is given by $2 \pi R_\mathrm{m} E = \pi R_\mathrm{m}^2 \mathrm{d}B_{\mathrm{app}}/\mathrm{d}t$ \cite{Caplin1994, Vanderbemden2002, Campbell1991}, i.e.
\begin{eqnarray} 
E = \frac{\mathrm{d}B_{\mathrm{app}}}{\mathrm{d}t}~\frac{R_\mathrm{m}}{2},
\label{e:Faraday}\end{eqnarray}
where $R_\mathrm{m}= a_1+e/2$ is the mean radius of the hollow cylinder.
As $B_{\mathrm{lim}}$ is proportional to the average current density, $J$, that flows in the superconductor, the relationship between $\mathrm{d}B_{\mathrm{app}}/\mathrm{d}t$ and $B_{\mathrm{lim}}$ corresponds, in a first approximation, to the $E$($J$) constitutive law. In the case of Bi-2212 tube, as well as for Bi-2223 \cite{Fagnard2009a, Fagnard2009b}, a power law fits adequately the $\mathrm{d}B_{\mathrm{app}}/\mathrm{d}t(B_\mathrm{lim})$  data obtained from such sweep creep experiments. Hence the experimental data can be used to determine the exponent of the $E$($J$) power law:
\begin{eqnarray} 
E(J) = E_{\mathrm{c}}~\left(\frac{J}{J_\mathrm{c}}\right)^n,
\label{e:EJ}\end{eqnarray}
where $E_\mathrm{c}$ is the electric field criterion defining the critical current density and $n$ is the creep exponent.
The procedure is performed at  temperatures between $10~\mathrm{K}$ and $77~\mathrm{K}$ and the temperature dependence of the power law exponent is presented in figure \ref{f:n_T}.

\subsection{Relaxation of the shielding currents} \label{ss:relaxation}

\begin{figure}
\center
\includegraphics[width=8cm]{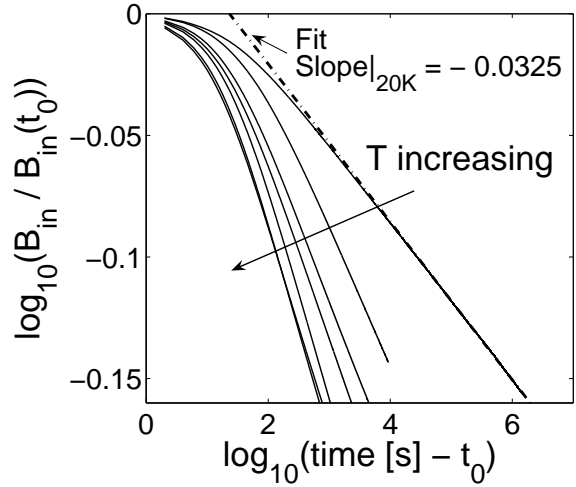}\caption
  {Relaxation of the remnant induction at the centre of the Bi-2212 hollow cylinder at $20~\mathrm{K}$  for 15 days and at $30~\mathrm{K}$, $40~\mathrm{K}$, $50~\mathrm{K}$, $60~\mathrm{K}$, $70~\mathrm{K}$, and $77~\mathrm{K}$ for $10000~\mathrm{s}$ after a $4~\mathrm{T}$ magnetization procedure. A power law whose exponent equals $-0.0325$ fits the experimental data at $20~\mathrm{K}$. This exponent can be related to the $E$($J$) exponent which gives $n = 31.8$ at $20~\mathrm{K}$ (see text). The \emph{n} values at the other temperatures are displayed in figure \ref{f:n_T}.}\label{f:creep0}
\end{figure}

The so-called \emph{relaxation creep experiments} are carried out at several fixed temperatures using the following procedure. First the Bi-2212 tube is subjected to an applied magnetic induction of $B_\mathrm{max}$, definitely large enough to bring the whole cylinder into the critical state. The applied magnetic field is then removed at a rapid constant sweep rate until it eventually goes to zero (at time $t_\mathrm{0}$) and a remnant magnetic moment is trapped. The value of $B_\mathrm{max}$ is set arbitrarily to $4~\mathrm{T}$ at all temperatures. Note that $4~\mathrm{T}$ corresponds to the value of the irreversibility field of our Bi-2212 material at $45~\mathrm{K}$. Above $45~\mathrm{K}$, the maximum magnetic induction of $4~\mathrm{T}$  exceeds the irreversible field $\mu_0$ $H_\mathrm{irr}$, but the final remnant state seems unaffected by this excursion in the reversible state. Similar results are observed by lowering the value of $B_\mathrm{max}$ with increasing temperature. after the field is removed, the decrease of the local magnetic induction at the centre of the cylinder, which is an image of the magnetization decrease in the superconducting wall, is recorded for $10^4~\mathrm{s}$ for experiments at $30$, $40$, $50$, $60$ and $77~\mathrm{K}$ and for $10^6~\mathrm{s}$ at $20~\mathrm{K}$. At each temperature, the magnetic induction at the centre of the cylinder, normalized by the value of $B_\mathrm{in}$ at $t_\mathrm{0}$, is plotted on a double logarithmic scaled graph in figure \ref{f:creep0}. After several tens of seconds, the decrease of the normalized inductions appears to follow a power law. The exponent can be related to the exponent of the $E$($J$) relationship. Indeed the time dependence of the remnant magnetization, $M$, can be expressed by a power law \cite{Caplin1994, Yamasaki2000, Anderson1962} whose index is given by
\begin{eqnarray} 
S = -\frac{\mathrm{d}\log(M)}{\mathrm{d}\log(t)}=\frac{1}{1-n}. 
\label{e:S}\end{eqnarray}

The exponent $S$ at $20~\mathrm{K}$ is determined to be equal $-0.0325$, which leads, using (\ref{e:S}), to a $n$ exponent of $31.8$ at $20~\mathrm{K}$. At $77~\mathrm{K}$, the value drops to $14$. Figure \ref{f:n_T} compares the \emph{n} values obtained from the relaxation creep experiments and those extracted from the analysis of the sweep creep measurements carried out at several temperatures.

A difference between sweep creep and relaxation creep experiments is that the electric field to be considered in relaxation measurements cannot be expressed by (\ref{e:Faraday}) as there is no varying applied magnetic field. Instead, it is related to the decrease of the shielding currents, which can be estimated by 
\begin{eqnarray} 
E = \frac{\mathrm{d}B_{\mathrm{in}}}{\mathrm{d}t}~\frac{R_\mathrm{m}}{2}. 
\label{e:Faraday2}\end{eqnarray}
In relaxation creep experiments, the accessible electric field range is one order of magnitude lower than for sweep creep experiments. 

\subsection{Results of AC shielding experiments} \label{ss:AC shielding results}

In AC shielding experiments, the applied magnetic field is produced by a primary coil and a secondary (pick-up) coil is used to probe the magnetic flux density $B_{\mathrm{in}}$ in the cylinder. The induced emf, $U(t)$, across the coil ($N_{\mathrm{pu}}$ turns, radius $R_{\mathrm{pu}}$) reads
\begin{eqnarray} 
U(t)=N_{\mathrm{pu}} \pi R_{\mathrm{pu}}^2 \frac{\mathrm{d}B_{\mathrm{in}}(t)}{\mathrm{d}t}.
\label{e:Upu}\end{eqnarray}
 In the case of perfect shielding, i. e. at small applied fields, one expects $B_{\mathrm{in}}=0$, which is the case for our Bi-2212 sample at $T=77~\mathrm{K}$: up to a primary current of $27~\mathrm{A_{rms}}$, no signal can be detected with the pick-up coil, i.e. shielding is complete within the limits of resolution. Above that value, a (non-sinusoidal) signal \cite{Elschner1999} is obtained which strongly increases with current.
From a numerical integration of $U(t)$, the peak value of the internal magnetic field, $B_\mathrm{in}$, can be calculated. In all our experiments this maximum field is smaller than $10~\%$ of the applied magnetic field. In a good approximation we therefore conclude that the mean current density in the superconductor is directly linked to the primary current, with 
\begin{eqnarray} 
J~e=\frac{N}{L} I_{p},
\label{e:mean current}\end{eqnarray}
where $L$ and $N$ denote the length and number of turns of the primary coil respectively.
This then also holds for the rms-values.
The electric field along the circumference at mean radius $R_m$ is also linked to the derivative of the magnetic field and is given in a good approximation (field variations within the material are neglected) by

\begin{eqnarray} 
E(t)~2 \pi R_m=\frac{\mathrm{d}B_{\mathrm{in}}(t)}{\mathrm{d}t}~\pi R_{\mathrm{m}}^2 = \frac{U_{\mathrm{pu}}(t)}{N_{\mathrm{pu}}~\pi R_{\mathrm{pu}}^2} \pi R_{\mathrm{m}}^2
\label{e:Einterm}\end{eqnarray}
or
\begin{eqnarray} 
E(t) = \frac{U_{\mathrm{pu}}(t) R_m}{2 N_{\mathrm{pu}} \pi R_{\mathrm{pu}}^2}.
\label{e:E-AC}\end{eqnarray}

The rms-value of the electric field is thus directly related to the rms-value of the non-sinusoidal pick-up-voltage. As a result, (\ref{e:mean current}), and (\ref{e:E-AC}) give us pairs ($J$, $E$) of rms-values. 

If the primary current is further slowly increased to values above $52~\mathrm{A}$, the observed pick-up signal abruptly increases by more than one order of magnitude and becomes sinusoidal \cite{Elschner1999}. This effect is due to an avalanche-type heating up of the sample with an at least partial quench.

Figure \ref{f:EJ} summarizes the results obtained from sweep creep, relaxation creep, transport and AC-shielding experiments used for the determination of the $E$($J$) law at $77~\mathrm{K}$.

\section{Discussion} \label{s:discussion}
\subsection{Temperature dependence of the critical current density}
\label{ss:temperature2}

As shown in figure \ref{f:Blim_T}, the $B_\mathrm{lim}$ value measured on the Bi-2212 tube increases monotonically with decreasing temperatures and reaches $1~\mathrm{T}$ at $10~\mathrm{K}$. To our knowledge, this value is largely in excess of the limit fields that have been reported so far for cuprate high-temperature superconductors. It gives also indirect evidence that Bi-2212 can sustain large magnetic forces that are due to the large magnetic flux density, in the presence of a high critical current density in the outer part of the cylinder. In the present Bi-2212 melt cast sample at $T=10~\mathrm{K}$, an order of magnitude of the stresses can be estimated by $\sigma \approx B^2/2\mu_0 \approx 0.4~\mathrm{MPa}$ \cite{Wilson1987} which is well below the fracture toughness of $25~\mathrm{MPa}$ (measured on the Bi-2212 cylinder at Nexans with a four-point-bending technique).  

According to the Bean model \cite{Bean1962}, the penetration field is proportional to the critical current density times the thickness of the superconductor. Both a high value of the critical current density and the thickness of the superconducting wall of the hollow cylinder are important to reach a high threshold value of the magnetic field that can be shielded.
Indeed, even if above $60~\mathrm{K}$ and at $\mathrm{d}B_{\mathrm{app}} /\mathrm{d}t = 125~\mathrm{\mu T/s}$, the critical current density of Bi-2212 is lower than that of the Bi-2223 material, as can be seen in figure \ref{f:compareJc}, the thickness difference leads to a higher value of $B_\mathrm{lim}$: for the Bi-2212 cylinder, $B_\mathrm{lim} = 21.9~\mathrm{mT}$, whereas for the Bi-2223 tube, we have $B_\mathrm{lim} = 13.7~\mathrm{mT}$. Below $60~\mathrm{K}$, the critical current density in the Bi-2212 material is now higher than the one of Bi-2223 and the Bi-2223 hollow cylinder is thinner than the Bi-2212 tube by a factor of 3.3. Then, as an example, the studied magnetic screen made of Bi-2212 material is able to shield a magnetic field 10 times higher than the Bi-2223 tube studied in \cite{Fagnard2009a} at $20~\mathrm{K}$. If one further considers that the current in the melt processed 2212 material is essentially carried within the partially textured zone of the cylinder wall \cite{Rikel2010}, the difference is even more impressive. 

Note that the magnetic field that can be shielded is determined by the product of the critical current density $J_\mathrm{c}$ and the wall thickness $e$. In the present case, the large $B_\mathrm{lim}$ values are obtained thanks to relatively ``thick'' wall of the shield, although the $J_\mathrm{c}$ itself is smaller than that reported usually for Y-123. It should be emphasized that magnetic shielding requires large current densities flowing on a macroscopic scale (i.e. the perimeter of the shield, typically several centimetres). For Y-123, this could be achieved with single domains or coated conductors with no joint resistance and good intergranular connectivity. The thickness of superconducting material in coated conductors is however much smaller (few microns) than that of bulk Bi-based materials but the high $J_\mathrm{c}$ in Y-123 may offer comparable shielding performances at $77~\mathrm{K}$ \cite{Levin2008, Fagnard2010a}. 

The  temperature dependence of the critical current density $J_\mathrm{c}$ can be described by the parametrisation 
\begin{eqnarray} 
J_\mathrm{c}(T) = J_\mathrm{c}(0)\left(1-\frac{T}{T_\mathrm{c}}\right)^\gamma
\label{e:JcT}\end{eqnarray}
used by Wesche \cite{Wesche1995} for Bi-2212/Ag wires and giving a value of $1.8$ for the exponent $\gamma$ for this material.
This law fits also, with a very good agreement, both $J_\mathrm{c}$($T$) data for Bi-2212 and Bi2223 cylinders, see Fig. \ref{f:compareJc}, and the parameters of the law in both cases are summarized in Table \ref{t:fitsJc}. 

\begin{table}
\caption{\label{t:fitsJc} $J_c$($T$) parameters of the Bi-2212 and the bi-2223 hollow cylinders.}
\begin{indented}
\item[]\begin{tabular}{@{}llll}
\br
Sample&$T_c$ ($K$)&$J_{c0}$ ($kA/cm^2$)&$\gamma$\\
\mr
Bi-2212&92&23&2.5\\
Bi-2223&108&5&1.5\\
\br
\end{tabular}
\end{indented}
\end{table}

For the Bi-2212 cylinder whose $T_\mathrm{c} = 92~\mathrm{K}$ \cite{Bock1990}, we find $J_\mathrm{c}(0) = 2.3~10^4~\mathrm{A/cm^2}$ and $\gamma = 2.5$ while for the Bi-2223 material whose $T_\mathrm{c} = 108~\mathrm{K}$ \cite{Denis2007a}, the parameters are $J_\mathrm{c}(0) = 5~10^3~\mathrm{A/cm^2}$ and $\gamma = 1.5$. The analytical expressions for the temperature dependence of $J_\mathrm{c}$ of these materials can be usefully employed in further numerical modelling to estimate the magnetic shielding properties at any temperature.

\subsection{Consequences of the non-linear $E$($J$) behaviour: sweep and relaxation creep effects}
\label{ss:EJconsequences}

We can observe in figure \ref{f:dBdt} that the threshold induction, $B_\mathrm{lim}$, above which the shielding is ineffective, is higher when the magnetic field is applied faster. In this case, the electric field, related to $\mathrm{d}B_{\mathrm{app}} /\mathrm{d}t$ by Faraday's law (\ref{e:Faraday}), is higher and allows stronger current density to flow in comparison with the measurements carried out at lower sweep rates. 

As can be seen in figure \ref{f:n_T}, in the temperature range  between $10~\mathrm{K}$ and $77~\mathrm{K}$, the \emph{n} exponents coming from sweep creep and relaxation creep experiments are in excellent agreement with one another even though they correspond to different electric field levels, as mentioned in section \ref{ss:relaxation}. These values of the \emph{n} exponent roughly agree with those reported in the literature on Bi-2212/Ag wires \cite{Wesche1995, Koizumi2003, Motowidlo2000, Kumakura2003, Hase2000, Kumakura2000} and on \emph{MCP} Bi-2212 material \cite{Cha2003}. In comparison, the $n$ values obtained on bulk Bi-2223 cylinder \cite{Fagnard2009a, Fagnard2009b} and on Bi-2223/Ag tapes \cite{Ayai2006} are generally higher than the one obtained on Bi-2212 compounds. For the Bi-2212 material, the small value of $n$ implies that there is a strong relaxation of the shielding currents that will make the magnetic screen less efficient for shielding magnetic fields close to the $B_\mathrm{lim}$ threshold. Associated to this shielding currents decay, the magnetic field will diffuse more easily than in Bi-2223 material whose $n$ exponent is higher.

\begin{figure}
\center
\includegraphics[width=8cm]{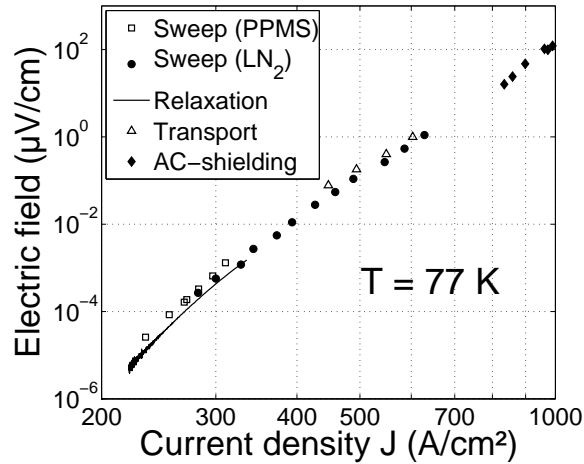}\caption {$E(J)$ characteristics of the Bi-2212 hollow cylinder at $T = 77~\mathrm{K}$. The data are obtained by (a) sweep creep experiment in PPMS [\opensquare], (b) sweep creep experiment [\fullcircle], (c) relaxation measurements [\full], (d) transport measurements [\opentriangle] and (e) AC-shielding experiments [\fulldiamond]. Techniques (b), (c), (d) and (e) are used in liquid nitrogen ($LN_2$).}\label{f:EJ}
\end{figure}

The $E$($J$) relationship can be figured out from various experiments (transport, AC-shielding, sweep creep, and relaxation creep) \cite{Miller1993, Caplin1994, Ries1992, Paul1991, Yamasaki2000, Fagnard2009b} which do not necessarily involve the same current paths: in transport experiments, the current flows along the axis of the cylinder whereas in magnetic experiments, shielding currents flow azimuthally. We believe that, because of the microstructure of the material produced by \emph{MCP}, we are allowed to present experimental data coming from different measurement techniques in the same graph. Indeed, as shown schematically in figure \ref{f:microstructure}, there is no influence of the microstructure in the inner and outer zones because they are untextured and the partially textured central area does not favour one of both experiments. In transport measurements, the currents flow in equal fractions of well-oriented grains (with \emph{c}-axis oriented in the azimuthal direction, $\mathbf{e_\theta}$, allowing the currents to flow in the ab-planes) and grains whose \emph{c}-axis are parallel to the direction of the superconducting currents, $\mathbf{e_z}$. For shielding measurements (AC-shielding, sweep creep, and relaxation creep), the \emph{c}-axis direction that is favourable to the current circulation is the axial one, $\mathbf{e_z}$, while another fraction is composed of badly oriented grains (along the azimuthal axis, $\mathbf{e_\theta}$). These considerations are mostly qualitative but justified by the symmetry with respect to the cooling gradient during the solidification process. 

Figure \ref{f:EJ} presents the results obtained from experiments used for the determination of the $E$($J$) law at $77~\mathrm{K}$. We can observe that the critical current density of $350~\mathrm{A/cm^2}$ reported in figure \ref{f:compareJc} for a given shielding factor of $1000$ and a sweep rate of $125~\mathrm{\mu T/s}$ corresponds according to (\ref{e:Faraday}) to a particular electric field of  $6.5~10^{-3}~\mathrm{\mu V/cm}$ and is consistent with the $E$($J$) data. Figure \ref{f:EJ} shows also that a continuous nearly ``linear" $E$($J$) curve is formed on a double logarithmic scaled graph, with a slope $n \approx 14$, confirming the power law behaviour in the range extending from $4~10^{-6}$ to $2~10^{2}~\mu\mathrm{V/cm}$. The exponent $n$ of the $E$($J$) power law is consistent with relaxation models such as flux-creep. Even at the highest measured electric fields, flux flow (\emph{FF}) or thermally assisted flux flow (\emph{TAFF}) regimes, characterized by a linear relationship between $E$ and $J$ \cite{Kes1989, Brandt1995} are not observed. However, for the lowest electric field levels, the $E$($J$) curve presents a slight downward curvature. This curvature has been reported on current-voltage characteristics measured on YBCO strips in \cite{Yamasaki2000} and on YBCO thin films by Kung et al. \cite{Kung1993} in accordance with the vortex-glass theory \cite{Koch1989} and the collective flux-creep theory \cite{Feigelman1989}. 

Investigating the $E$($J$) law in regions of smaller electric fields would require observation times of one hundred of days to gain one decade of voltage. An extension of $E$($J$) to higher electric fields is, due to heating, only possible with pulsed current transport techniques \cite{Sirois2010}. At the required high currents this is difficult for our macroscopic samples but pulse current experiments were carried out on similar MCP samples of smaller size. It should be emphasized that our $E$($J$) measurements on large tubular Bi-2212 samples are entirely consistent with those reported up to $0.1~\mathrm{V/cm}$ on small rods of MCP Bi-2212 \cite{Elschner2001}.

\section{Conclusions} \label{s:conclusion}

In this paper, we have shown that Bi-2212 hollow cylinders, produced by the \emph{MCP} technique, could be used as very efficient magnetic shields, especially at temperature below $60~\mathrm{K}$ where the critical current exceeds the critical current of Bi-2223 polycrystalline tubes. The temperature dependence of the average critical current has been determined from the magnetic shielding experiments. We have shown that, at $T=10~\mathrm{K}$, the studied Bi-2212 sampled is able to shield a record value of $1~\mathrm{T}$ magnetic flux density.

Next, the influence of the non-linear $E$($J$) relationship on the shielding properties has been investigated by various experiments. First, the effect of the sweep rate of the applied magnetic field, $\mathrm{d}B_{\mathrm{app}} /\mathrm{d}t$, on the threshold magnetic induction below which the shielding occurs, $B_{\mathrm{lim}}$, has been studied at several temperatures below $77~\mathrm{K}$. At a given temperature, the $\mathrm{d}B_{\mathrm{app}}/\mathrm{d}t (B_{\mathrm{lim}})$ data show a power law dependence whose exponent is related to the $n$ exponent of the $E$($J$) power law. A second set of experiments was carried out in order to measure the shielding currents relaxation and determine the $n$ exponent of the $E$($J$) law at lower electric fields than those attainable in sweep creep experiments.

A piecewise $E$($J$) graph was then assembled combining sweep creep and relaxation creep experimental results as well as direct transport and AC-shielding measurements, all experiments showing a good mutual agreement. At $77~\mathrm{K}$, a power law with a single exponent was confirmed by all these experimental techniques on a wide electric field range but a slight downward curvature of the $\log E$($\log J$) law was observed at very low electric fields.

The results presented in this paper have shown that melt cast Bi-2212 hollow cylinders are extremely promising to shield large magnetic fields at low temperature. The centrifugal melt casting process  is a very flexible technique that allows shields to be manufactured with a wide range of axisymmetric geometries and dimensions of samples. The shields can be easily machined in order to assemble several cylinders to obtain magnetic shields of larger dimensions while avoiding excessive magnetic flux leakage. 


\section*{Acknowledgments}

The authors thank the Royal Military Academy, FNRS and ULg for cryofluid, equipment and travel grants.
We thank Mark Rikel, Nexans Superconductors, for valuable advice concerning the microstructural issues and a critical reading of the manuscript.



\section*{References}
\begin{thebibliography}{10}

\bibitem{Herrmann1993} Herrmann P, Albrecht C, Bock, J, Cottevieille C, Elschner S, Herkert W, Lafon M-O, Lauvray H, Leriche A, Nick W, Preisler E, Salzburger H, Tourre J.-M. and Verhaege T 1993 \emph{IEEE Trans. Appl. Supercond.} \textbf{3} 876 

\bibitem{Dommerque2010} Dommerque R, Kr\"{a}mer S, Hobl A, B\"{o}hm R, Bludau M, Bock J, Klaus D, Piereder H, Wilson A, Kr\"{u}ger T, Pfeiffer G, Pfeiffer K and Elschner S 2010 \emph{Supercond. Sci. Technol.} \textbf{23} 034020

\bibitem{Grohmann1974} Grohmann K, Hahlbohm H, L\"{u}bbig H and Ramin H 1974 \emph{Cryogenics} \textbf{14} 499 

\bibitem{Giunchi2007} Giunchi G, Bassani E, Cavallin T, Bancone N and Pavese F 2007 \emph{Supercond. Sci. Technol.} \textbf{20} L39 

\bibitem{Ohta1999} Ohta H, Aono M, Matsui T, Uchikawa Y, Kobayashi K, Tanabe K, Takeuchi S, Narasaki K, Tsunematsu S, Koyabu Y, Kamekawa Y, Nakayama K, Shimizu T, Koike K, Hoshino K, Kotaka H, Sudoh E, Takahara H, Yoshida Y, Shinada K, Takahata M, Yamada Y and Kamijo K 1999 \emph{IEEE Trans. Appl. Supercond.} \textbf{9} 4073

\bibitem{Denis2007a} Denis S, Dusoulier L, Dirickx M, Vanderbemden P, Cloots R, Ausloos M and Vanderheyden B 2007 \emph{Supercond. Sci. Technol.} \textbf{20} 192

\bibitem{Fagnard2009a} Fagnard J-F, Dirickx M, Ausloos M, Lousberg G, Vanderheyden B and Vanderbemden P. 2009 \emph{Supercond. Sci. Technol.} \textbf{22} 105002

\bibitem{Bock1995} Bock J, Elschner S and Herrmann P 1995 \emph{IEEE Trans. Appl. Supercond.} \textbf{5} 1409 

\bibitem{Muller1999} M\"{u}ller K-H , Andrikidis C, Du J, Leslie K E and Foley C. P 1999 \emph{Phys. Rev. B} \textbf{60} 659 

\bibitem{Fagnard2002} Fagnard J-F, Vanderbemden P, Crate D, Misson V, Ausloos M and Cloots R 2002 \emph{Physica C} \textbf{372} 970

\bibitem{Miller1993} Miller M M, Caroll T, Soulen R, Toth L, Rayne R, Alford N McN and Saunders C S 1993 \emph{Cryogenics} \textbf{33} 180

\bibitem{Plechacek1996} Plech\'{a}cek V, Pollert E and Hejtm\'{a}nek J 1996 \emph{Mater. Chem. Phys.} \textbf{43} 95

\bibitem{Cavallin2006} Cavallin T, Quarantiello R, Matrone A and Giunchi G 2006 \emph{J. Phys. Conf. Ser.} \textbf{43} 1015

\bibitem{Fournier1994} Fournier P and Aubin M 1994 \emph{Phys. Rev. B} \textbf{49} 15976

\bibitem{Caplin1994} Caplin A, Cohen L, Perkins G and Zhukov A 1994 \emph{Supercond. Sci. Technol.} \textbf{7} 412

\bibitem{Vanderbemden2002} Vanderbemden P, Misson V, Ausloos M and Cloots R 2002 \emph{Physica C} \textbf{372} 1225

\bibitem{Ries1992} Ries G, Neumuller H. W and Schmidt W 1992 \emph{Supercond. Sci. Technol.} \textbf{5} S81

\bibitem{Paul1991} Paul W, Hu D and Baumann T 1991 \emph{Physica C} \textbf{185-189} 2373

\bibitem{Yamasaki2000} Yamasaki H and Mawatari Y 2000 \emph{Supercond. Sci. Technol.} \textbf{13} 202

\bibitem{Denis2007b} Denis S,  Dirickx M, Vanderbemden P, Ausloos M and Vanderheyden B 2007 \emph{Supercond. Sci. Technol.} \textbf{20} 418

\bibitem{Fagnard2009b} Fagnard J-F, Denis S, Lousberg G, Dirickx M, Ausloos M, Vanderheyden B and Vanderbemden P 2009 \emph{IEEE Trans. Appl. Supercond.} \textbf{19} 2905

\bibitem{Rikel2010} Rikel M O, Elschner S, Hobl A, Hasenh\"{u}tl A, Klein M, Bock J \emph{IEEE Trans. Appl. Supercond.} (Proc. ASC 2010), submitted

\bibitem{Elschner1992} Elschner S, Bock J 1992 \emph{Adv. Mat.} \textbf{4} 242

\bibitem{Elschner1999} Elschner S, Bock J Brommer G, Cowey L 1999 \emph{Appl. Supercond.} \textbf{167} 1029

\bibitem{Bean1962} Bean C 1962 \emph{Phys. Rev. Lett.} \textbf{8} 250

\bibitem{Campbell1991} Campbell A M 1991 \emph{Magnetic susceptibility of superconductors and other spin systems} (New-York: Plenum Press) pp 129-155

\bibitem{Anderson1962} Anderson P W 1962 \emph{Phys. Rev. Lett.} \textbf{9} 309

\bibitem{Wilson1987} Wilson M 1987 \emph{Superconducting Magnets} (Oxford: Oxford University Press)

\bibitem{Levin2008} Levin G A, Barnes P N, Murphy J, Brunke L, Long J D, Horwath J and Turgut Z 2008 \emph{Appl. Phys. Lett.} \textbf{93} 062504

\bibitem{Fagnard2010a} Fagnard J-F, Dirickx M, Levin G A, Barnes P N, Vanderheyden B and Vanderbemden P 2010 \emph{J. Appl. Phys.} \textbf{108} 013910

\bibitem{Wesche1995} Wesche R 1995 \emph{Physica C} \textbf{246} 186

\bibitem{Bock1990} Bock J, Preisler E and Elschner S 1990 \emph{Proc. Int. Conf. ICMAS 90} (R. Tournier and R. Suryanarayanan, eds)

\bibitem{Koizumi2003} Koizumi T, Nakatsu T, Ohtani N, Aoki Y, Hasegawa T, Hirano N and Nagaya S 2003 \emph{Physica C} \textbf{392-396} 1025

\bibitem{Motowidlo2000} Motowidlo L R, Selvamanickam V, Galinski G, Vo N, Haldar P and Sokolowski R. S 2000 \emph{Physica C} \textbf{335} 44

\bibitem{Kumakura2003} Kumakura H, Matsumoto A, Sung Y. S and Kitaguchi H 2003 \emph{Physica C} \textbf{384} 283

\bibitem{Hase2000} Hase T, Murakami Y, Hayashi S, Kawate Y, Kiyoshi T, Wada H, Sairote S and Ogawa R 2000 \emph{Physica C} \textbf{335} 6

\bibitem{Kumakura2000} Kumakura H, Kitaguchi H, Miao H, Togano K, Koizumi T and Hasegawa T 2000 \emph{Physica C} \textbf{335} 31

\bibitem{Cha2003} Cha Y S 2003 \emph{IEEE Trans. Appl. Supercond.} \textbf{13} 2028

\bibitem{Ayai2006} Ayai N, Kato T, Fujikami J, Fujino K, Kobayashi S, Ueno E, Yamazaki K, Kikuchi M, Ohkura K, Hayashi K, Sato K and Hata R 2006 \emph{J. Phys.: Conf. Ser.} \textbf{43} 47

\bibitem{Kes1989} Kes P H, Aarts J, van den Berg J, van der Beek C J and Mydosh J A 1989 \emph{Supercond. Sci. Technol.} \textbf{1} 242

\bibitem{Brandt1995} Brandt E H 1995 \emph{Rep. Prog. Phys.} \textbf{58} 1465

\bibitem{Kung1993} Kung P J, Maley M P, McHenry M E, Willis J O, Murakami M and Tanaka S 1993 \emph{Phys. Rev. B} \textbf{48} 13922

\bibitem{Koch1989} Koch R H, Foglietti V, Gallagher W J, Koren G, Gupta A and Fisher M P A 1989 \emph{Phys. Rev. Lett.} \textbf{63} 1511

\bibitem{Feigelman1989} Feigel'man M V, Geshkenbein V B, Larkin A I and Vinokur V M 1989 \emph{Phys. Rev. Lett.} \textbf{63} 2303

\bibitem{Sirois2010} Sirois F, Coulombe J, Roy F and Dutoit B 2010 \emph{Supercond. Sci. Technol.} \textbf{23} 034018

\bibitem{Elschner2001} Elschner S, Breuer F, Wolf A, Noe M, Cowey L and Bock J 2001 \emph{IEEE
Trans. Appl. Supercond.} \textbf{11} 2507

\end{thebibliography}
\end{document}